\documentclass{iau}
\usepackage{graphicx}

\title[Rapid variations in SuperWASP] 
{Rapidly varying A-type stars in the SuperWASP archive}

\author[Daniel L. Holdsworth \& Barry Smalley]   
{Daniel L. Holdsworth
 \and Barry Smalley}

\affiliation{Astrophysics Group, Keele University, Staffordshire, ST5 5BG, United Kingdom
\\ email: {\tt d.l.holdsworth@keele.ac.uk} \\[\affilskip]
}

\pubyear{2013}
\volume{301}  
\pagerange{xxx--xxx}
\jname{Precision Asteroseismology}
\editors{W. Chaplin, J. Guzik, G. Handler \& A. Pigulski}
\begin{document}

\maketitle

\begin{abstract}
The searches for transiting exoplanets have produced a vast amount of time-resolved photometric data of many millions of stars. One of the leading ground-based surveys is the SuperWASP project. We present the initial results of a survey of over $1.5$~million A-type stars in the search for high frequency pulsations using SuperWASP photometry. We are able to detect pulsations down to the $0.5$~mmag level in the broad-band photometry. This has enabled the discovery of several rapidly oscillating Ap stars and over $200$~delta Scuti stars with frequencies above $50$~d$^{-1}$, and at least one pulsating sdB star. Such a large number of results allows us to statistically study the frequency overlap between roAp and delta Scuti stars and probe to higher frequency regimes with existing data.

\end{abstract}

\firstsection 
\section{SuperWASP}

The Wide Angle Search for Planets (WASP) is a two site campaign in the search for transiting exoplanets (\cite[Pollacco et. al 2006]{Pollacco06}). Each observatory consists of $8$~telephoto lenses mounted in a $2\times4$ configuration. To date there are over $31$~million objects in the WASP archive. Observations consist of two consecutive $30$s integrations followed by a $10$~minute gap. The entire observable sky can be visited every $40$~minutes. The short integrations, and non-uniform sampling, allow for a Nyquist frequency of up to $1440$~d$^{-1}$.

Due to the observing strategy, observations of a single star can occur over many seasons, with the target sometimes appearing in more than one camera. This provides a long time-base of observations which can either be combined or split into individual data sets.

\section{Methodology}
We selected, using 2MASS colours, over $1.5$~million A-type and earlier stars from the SuperWASP archive. We aimed to identify new pulsating systems with frequencies above $50$~d$^{-1}$. Such a frequency range allows for the discovery of new $\delta$~Scuti stars, rapidly oscillating Ap (roAp) stars and compact pulsators. Thus, enabling a statistical study on potential frequency boundaries.

We calculated a periodogram for each WASP season for every individual object -- over $9$~million lightcurves. Periodograms for an object are cross-checked for peaks at the same frequency. An object is required to have two or more concurrent peaks to be thought of as genuine. Spectroscopic follow-up is sought for the most convincing candidates. For our northern targets we make use of the service mode on WHT/ISIS and UK SALT Consortium (UKSC) time on SALT/RSS for our southern targets.

\section{Results}

{\bf{roAp Stars}}

\begin{figure}[b]
\begin{center}
 \includegraphics[width=50mm,angle=90,trim=0mm 5mm 0mm 0mm, clip]{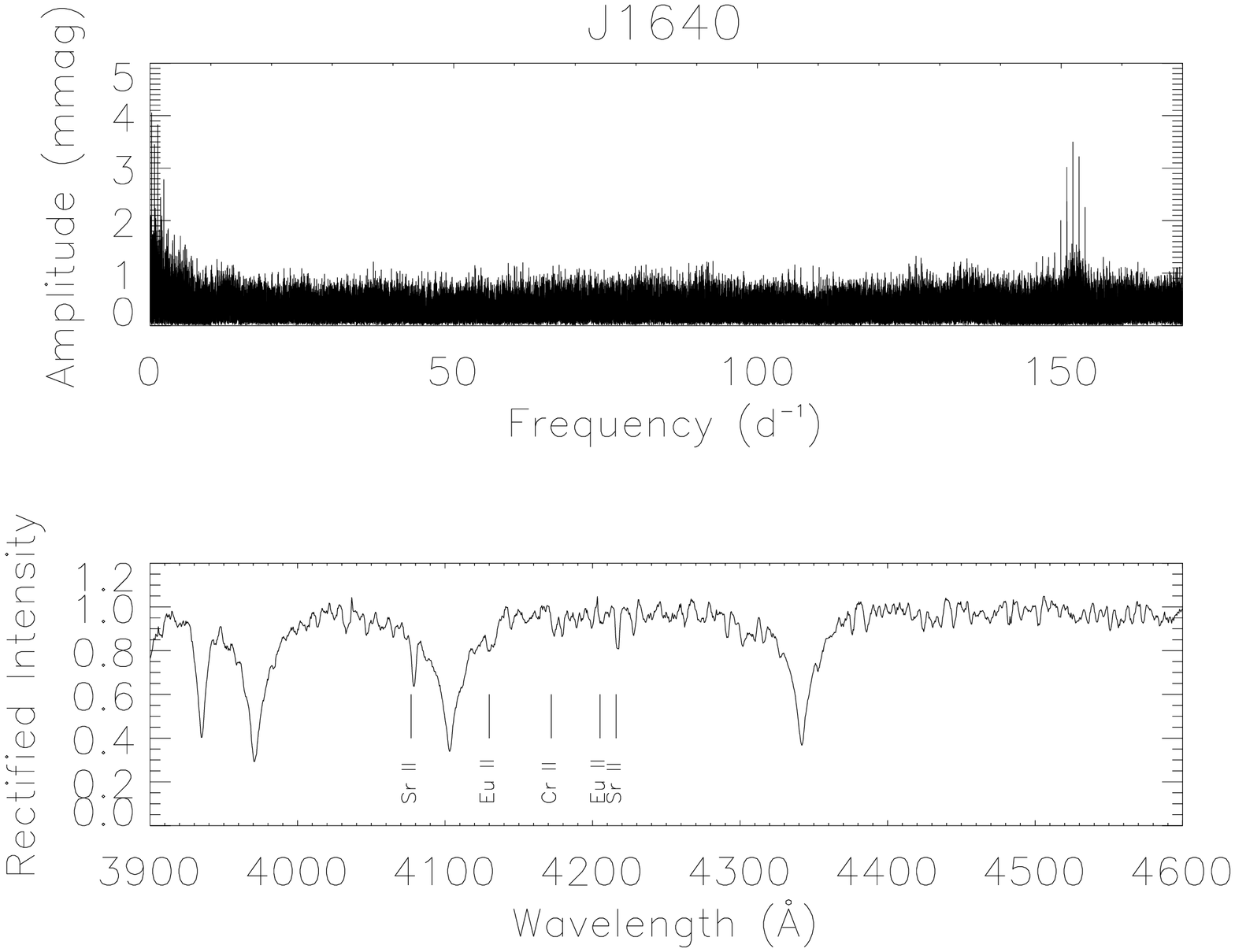} 
 \includegraphics[width=50mm,angle=90,trim=0mm 5mm 0mm 0mm, clip]{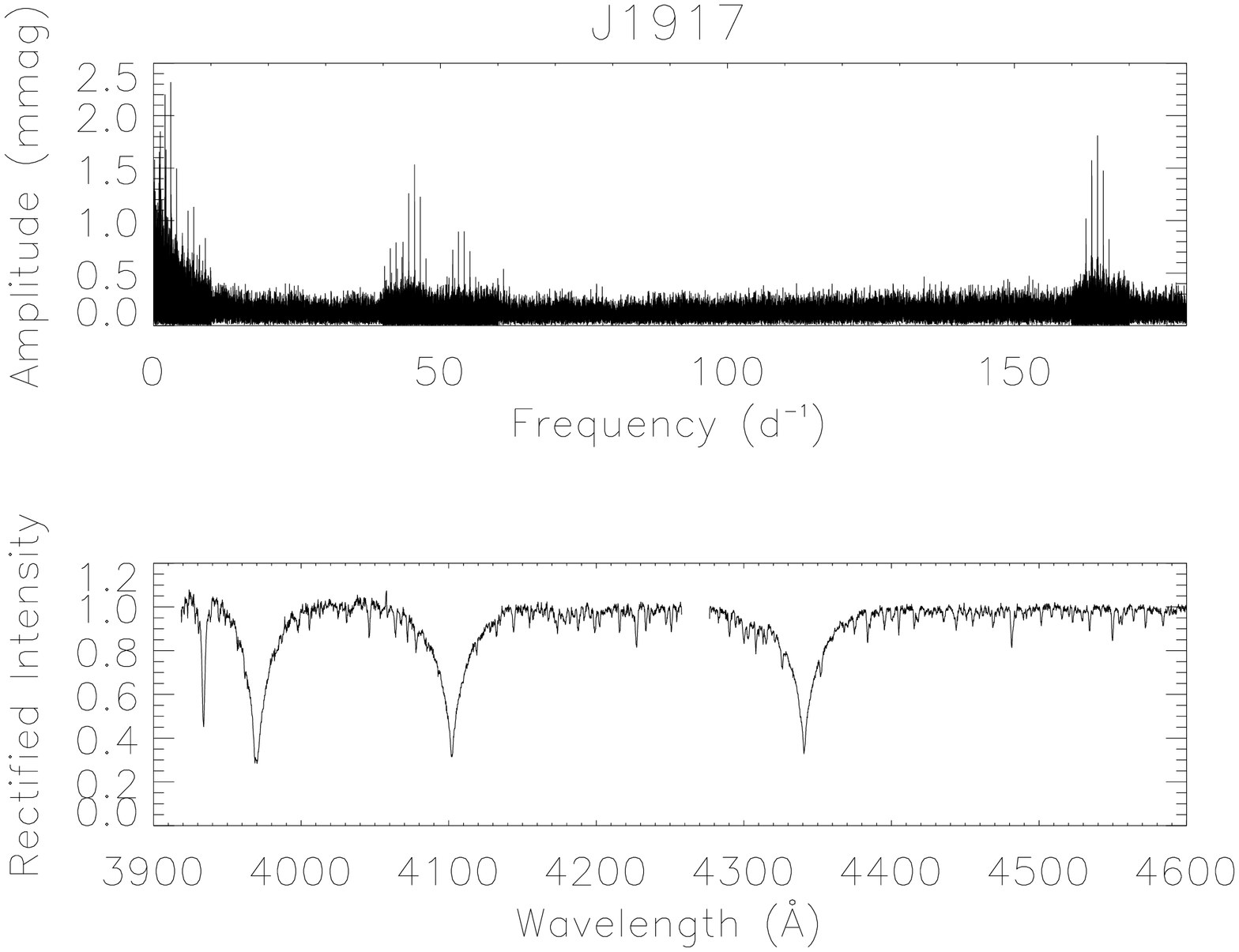}
 \caption{{\textit{Left:}} An example of a WASP roAp periodogram and WHT/ISIS spectrum confirming the Ap nature of the star. {\textit{Right}} A hybrid target showing both roAp and $\delta$~Scuti pulsations. The SALT/RSS spectrum shows this is an Am star due to the weak Ca K line.}
   \label{fig1}
\end{center}
\end{figure}

To identify candidate roAp stars we require a target to show a single peak in the periodogram above $50$~d$^{-1}$. This peak must appear in more than one season of WASP data to eliminate the possibility of spurious peaks due to noise. Using this criteria we identify $21$~candidate stars. Classification spectra have been obtained for these targets, confirming at least $10$~new roAp stars (e.g. Fig.\,\ref{fig1} {\textit{Left}}).\\

{\bf{$\delta$ Scuti Stars}}

Candidate $\delta$~Scuti stars are identified as those objects with multiple frequencies $>50$~d$^{-1}$ and having their principle peak in multiple seasons. We have $>200$~candidates up to a maximum frequency of $100$~d$^{-1}$. Spectra have been secured for photometrically interesting targets such as J$1917$~(Fig.\,\ref{fig1} {\textit{Right}}). $52\%$ of the targets for which we currently have spectra are pulsating Am stars (also see \cite[Smalley \etal\ 2011]{smalley11}).

\begin{table}
  \begin{center}
  \caption{Current statistics from the WASP survey.}
  \label{tab1}
 {\scriptsize
  \begin{tabular}{|c|c|c|c|}\hline 
{\bf Pulsation Class} & {\bf Candidates} & {\bf Number with Spectra} & {\bf{Confirmed}} \\ 
roAp & $21$ & $21$ & $10$ \\
$\delta$~Scuti (Am) & $204$ & $18^{*}$ & $29^{*}$($15$)\\ \hline
  \end{tabular}\\
$^{*}$Some roAp candidates are found to be $\delta$~Scuti, hence the discrepancy.
  }
 \end{center}
\end{table}

\section{Conclusions}

The SuperWASP archive has provided a new approach in searching for stellar pulsations in the A-type stars. We have thus far identified $10$~new roAp targets for further in depth study as well as highlighting over $200$~potential new $\delta$~Scuti stars with periods shorter than $\sim30$~minutes. The archive holds many more examples of pulsating systems which are yet to be exploited.

\end{document}